\documentclass[sigconf]{acmart}
\usepackage{booktabs} 
\usepackage[nolist,nohyperlinks]{acronym}
\usepackage{multirow}
\setcopyright{rightsretained}
\usepackage{algorithm}
\usepackage{algorithmic}
\usepackage[skip=0.4\baselineskip]{caption}
\usepackage{balance}

\makeatletter
\def\hlinewd#1{%
  \noalign{\ifnum0=`}\fi\hrule \@height #1 \futurelet
   \reserved@a\@xhline}
\makeatother

\acmDOI{10.475/123_4}

\acmISBN{123-4567-24-567/08/06}

\acmConference[SIGIR'19]{ACM sigir conference}{July 21-25, 2019}{Paris, France}
\acmYear{2019}
\copyrightyear{2019}

\acmArticle{4}
\acmPrice{15.00}

\renewcommand\footnotetextcopyrightpermission[1]{}
\begin{document}
\title{Improving Collaborative Metric Learning with Efficient Negative Sampling}
\author{Viet-Anh Tran, Romain Hennequin, Jimena Royo-Letelier, Manuel Moussallam}
\affiliation{%
  \institution{Deezer Research \& Development, Paris, France}
  \streetaddress{}
  \city{}
  \state{}
  \postcode{}
}

\email{research@deezer.com}




\renewcommand{\shortauthors}{Viet-Anh Tran, Romain Hennequin, Jimena Royo-Letelier, Manuel Moussallam}

\begin{acronym}
    \acro{CML}{Collaborative Metric Learning}
    \acro{MF}{Matrix Factorization}
    \acro{MAP}{Mean Average Precision}
    \acro{NDCG}{Normalized Discounted Cumulative Gain}
    \acro{GOR}{Global Orthogonal Regularization}
    \acro{MSD}{Million Song Dataset}
    \acro{MMR}{Mean of Median Rank}
    \acro{CF}{Collaborative Filtering}
\end{acronym}

\begin{abstract}
Distance metric learning based on triplet loss has been applied with success in a wide range of applications such as face recognition, image retrieval, speaker change detection and recently recommendation with the \ac{CML} model. However, as we show in this article, CML requires large batches to work reasonably well because of a too simplistic uniform negative sampling strategy for selecting triplets. Due to memory limitations, this makes it difficult to scale in high-dimensional scenarios. To alleviate this problem, we propose here a 2-stage negative sampling strategy which finds triplets that are highly informative for learning. Our strategy allows CML to work effectively in terms of accuracy and popularity bias, even when the batch size is an order of magnitude smaller than what would be needed with the default uniform sampling. We demonstrate the suitability of the proposed strategy for recommendation and exhibit consistent positive results across various datasets.

\end{abstract}

%
%



\copyrightyear{2019} 
\acmYear{2019} 
\setcopyright{acmlicensed}
\acmConference[SIGIR '19]{Proceedings of the 42nd International ACM SIGIR Conference on Research and Development in Information Retrieval}{July 21--25, 2019}{Paris, France}
\acmBooktitle{Proceedings of the 42nd International ACM SIGIR Conference on Research and Development in Information Retrieval (SIGIR '19), July 21--25, 2019, Paris, France}
\acmPrice{15.00}
\acmDOI{10.1145/3331184.3331337}
\acmISBN{978-1-4503-6172-9/19/07}

\maketitle

\section{Introduction}
Distance metric learning aims at representing data points in a space where proximity accounts for similarity. A recent popular approach in face recognition \cite{schroff:cvpr15}, image retrieval \cite{song:cvpr16} or speaker change detection \cite{bredin:icassp17} formalizes this problem as a triplet loss optimization task, namely minimizing: $L = max(D(a,p) - D(a,n) + \alpha, 0)$ where $D(a,p)$ is the distance between \textit{intra-class} (same label) samples (anchor and positive), $D(a,n)$ is the distance between \textit{inter-class} (different labels) samples (anchor and negative) and $\alpha > 0$ is a margin constant. The main idea is to enforce inter-class pairs to be far away from intra-class pairs at least by a margin $\alpha$. This favors clustering of same class samples. As pointed out in \cite{wang:iccv17, hermans:arxiv17}, minimizing $L$ is not easy as the number of possible triplets grows cubically with the number of identities. Furthermore, a naive uniform sampling strategy would select trivial triplets for which the gradient of $L$ is negligible. As a result, learning may be slow and stuck in a local minima \cite{wu:iccv17}. To address this problem, some works proposed to select only \textit{hard} samples ($D(a,p) > D(a,n)$) for training \cite{simoserra:iccv15, shrivastava:cvpr16}. Hard samples mining, however, selects triplets with noisy (high variance) gradients of $L$. Models may then struggle to effectively push inter-class pairs apart, and end up in a collapsed state \cite{schroff:cvpr15, wu:iccv17}. A relaxed alternative is to mine only \textit{semi-hard} samples \cite{schroff:cvpr15}: triplets in which the negative is not necessarily closer to the anchor than the positive, but which still produce a strictly positive loss. This strategy improves the robustness of training by avoiding overfitting outliers in the training set \cite{harwood:iccv2017}. It typically converges quickly in the first iterations, but eventually runs out of informative samples and stops making progress. In \cite{wu:iccv17} authors attributed this phenomenon to the concentration of the gradient's variance of $L$ for semi-hard samples to a small region. To address this issue, they proposed to select negative samples based on their distances to anchors. They demonstrated that this strategy results in the variance of the gradient of $L$ being spread in a larger range, and thus consistently produces informative triplets \cite{wu:iccv17}.

Its ability to deal with large-scale catalogs and data sparsity \cite{wang:iccv17} makes the triplet loss model suitable for recommendation tasks. It has indeed been recently proposed as the \ac{CML} model \cite{hsieh:www17}, reaching competitive results with traditional \ac{MF} methods \cite{yifan:icdm08, rendle:auai09}. \ac{CML} assumes that users and items can be placed in a joint low dimensional metric-space. Recommendations are then easily done based on their proximity measured by their Euclidean distance. \ac{CML} can achieve competitive accuracy \cite{hsieh:www17} but we show in this paper that it requires large batches to do so, because of it's simplistic uniform negative sampling strategy. Owing to memory limitations, this makes \ac{CML} unable to scale in high-dimensional scenarios, \emph{e.g.}, when building a hybrid multimedia recommender system that learns jointly from interaction data and high-dimensional item contents such as audio spectrograms \cite{lee:arxiv18}. For that reason, following the idea in \cite{wu:iccv17}, we replace the default uniform sampling by a 2-stage strategy, which finds triplets that are consistently informative for learning. This enables \ac{CML} to be competitive with uniform sampling, even with small batches, both in terms of accuracy and popularity bias.

Our contributions are threefold: (1) We study the influence of batch size on the CML's performance. (2) We propose a 2-stage negative sampling that makes CML efficient with small batches. (3) We demonstrate the suitability of our sampling strategy on three real-world datasets, for the Top-N recommendation task, in terms of accuracy and popularity bias. We note especially a significant improvement over standard \ac{CML} on music recommendation. We also provide code to reproduce our results\footnote{Code available at: \href{https://github.com/deezer/sigir2019-2stagesampling}{https://github.com/deezer/sigir2019-2stagesampling}}.
\section{Preliminaries}
\subsection{Problem Formulation}
Consider a dataset with $N$ users, $M$ items and the binary interaction ${M \times N}$ matrix $R$,  where $R_{ij}$ indicates the only positive implicit feedback (\emph{e.g.}, clicks, listens, view histories logs etc.) between the $i$-th user and the $j$-th item. We use $\mathbf{S} = \{(i, j)\text{ }|\text{ }R_{ij} = 1\}$ to denote the set of user-item pairs where there exists implicit interactions. The considered task is to predict the items/users that are likely to interact together.

\subsection{Collaborative Metric Learning}
\ac{CML} \cite{hsieh:www17} learns a joint metric space of users and items to encode $\mathbf{S}$. The idea is to learn a metric that pulls the positive pairs in $\mathbf{S}$ closer while pushing the negative pairs (pairs not in $\mathbf{S}$) relatively further apart compared to the positive ones, based on the following loss:
\begin{equation}
\begin{split}
    L^{\text{triplet}} = \sum_{(i,j) \in B}w_{ij}[D^2(\mathbf{u}_{i},\mathbf{v}_{j})-\min_{k \in N_{ij}}D^2(\mathbf{u}_{i}, \mathbf{v}_{k}) + \alpha]_{+} + \lambda_c L_c\\
    \text{s.t. } \forall p \leq M, q \leq N: ||\mathbf{u}_p||_2 \leq 1, ||\mathbf{v}_{q}||_2 \leq 1
\end{split}
\end{equation}
where $\mathbf{u}_{i}$, $\mathbf{v}_{j}$ are, respectively, user and item latent vectors in $\mathbb R^d$, $B \subset S$ is the set of positive pairs in the considered mini-batch, $N_{ij} \subset \{k | (i, k) \not \in S \}$ is a set of negative samples per triplet, $\alpha > 0$ is a margin constant, $D$ is the Euclidean distance and $w_{ij}$ is a weight based on the number of negatives in $N_{ij}$ falling inside the $\alpha$-ball to penalize items at a lower rank \cite{weston:ml10}, $[.]_+ = max(., 0)$, $L_c$ is regularization term (weighted by the hyper parameter $\lambda_c$) used to de-correlate the dimensions in the learned metric \cite{hsieh:www17}. The recommendation for an user is then made by finding the $k$ nearest items around her/him in the latent space.

In this work, we set $w_{ij}$ to 1 for fair comparison between different sampling strategies. Furthermore, we do not use $L_c$ for all models because we have inferior results for the uniform sampling with this regularization (with the code provided by authors on github\footnote{Original CML \href{https://github.com/changun/CollMetric}{https://github.com/changun/CollMetric}}). Additionally, all user and item vectors are normalized to the unit sphere: $\forall p \leq M, q \leq N: ||\mathbf{u}_p||_2 = 1, ||\mathbf{v}_{q}||_2 = 1$ (by adding a $L_2$-normalization step after the user/item embedding layer) instead of being bound within the unit ball.

\section{Sampling Strategy}
\subsection{Spread-out Regularization}
In \cite{zhang:iccv17}, the authors argued that in order to fully exploit the expressive power of the embedding, latent vectors should be sufficiently "spread-out" over the space. Intuitively, two randomly sampled non-matching vectors are "spread-out" if they are orthogonal with high probability. To this end, they proved that if $\mathbf{p}_{1}, \mathbf{p}_{2}$ are two vectors independently and uniformly sampled from the unit sphere in $\mathbb{R}^d$, the probability density of $\mathbf{p}_{1}^{T}\mathbf{p}_{2}$ satisfies
  \begin{displaymath}
    p(\mathbf{p}_{1}^{T}\mathbf{p}_{2} = s) = \begin{cases}
        \frac{(1 - s^{2})^{\frac{d-1}{2}-1}}{\text{Beta}(\frac{d-1}{2}, \frac{1}{2})} & \text{if} \, -1 \leq s \leq 1 \\
    0 & \text{otherwise}
    \end{cases}
  \end{displaymath}
  where $\text{Beta}(a, b)$ is the beta distribution function.
From this distribution, they further found that $\mathbb{E}\left[\mathbf{p}_{1}^{T}\mathbf{p}_{2}\right] = 0 \text{ and } \mathbb{E}\left[(\mathbf{p}_{1}^{T}\mathbf{p}_{2})^2\right] = \frac{1}{d}$, and proposed the \ac{GOR} to enforce the spread of latent vectors. The application of \ac{GOR} for \ac{CML} is thus:


\begin{align}
    L &= L^{\text{triplet}} + \lambda_g L^{\text{GOR}} \\
    L^{\text{GOR}} &= \Big(\frac{1}{Q}\sum_{(i,j) \in B}\sum_{k \in N_{ij}}\mathbf{v}_{j}^{T}\mathbf{v}_{k}\Big)^{2} + \Big[ \frac{1}{Q}\sum_{(i,j) \in B}\sum_{k \in N_{ij}}(\mathbf{v}_{j}^{T}\mathbf{v}_{k})^2 - \frac{1}{d}\Big]_+
\end{align}
where $\lambda_g$ is an hyperparameter, $Q = |B| \times |N_{ij}|$ and $d$ is the dimension of the latent space.
\subsection{2-stage negative sampling}
\label{2_stage_negative_sampling}
To construct a batch, we first randomly sample pairs in $S$ as in \cite{hsieh:www17} to get the anchor users and the positive items. Our strategy aims at replacing the uniform sampling for the set $N_{ij}$ negative items in a triplet by a 2-stage setting as described below.

In the first stage, we sample $C$ negative candidates from all items in the dataset based on their frequencies as proposed in the popular Word2Vec algorithm in natural language processing \cite{mikolov:nips12} and its application for the recommendation task \cite{barkan:mlsp2016, dupre:recsys17, musto:recsys15}: 
\begin{equation}
\label{w2vsampling}
    \Pr(j)=\frac{f(j)^\beta}{\sum_{j'}f(j')^\beta} \,,
\end{equation}

\noindent where $f(j)$ is the interaction frequency of item $j$ and the parameter $\beta$ plays a role in sharpening or smoothing the distribution. A positive $\beta$ leads to a sampling that favors popular items, a $\beta$ equal to 0 leads to items being sampled uniformly, while a negative $\beta$ makes unpopular items being more likely sampled. In this work, we use a positive $\beta$ to favor popular items as negative samples. The motivation is that due to the popularity bias in interaction data \cite{steck:recsys11}, popular items tend to be close together. A challenge is thus to push non-matching popular items farther away in the latent space. Spreading popular items apart could then help to reduce the popularity bias often witnessed in recommendation.

In the second stage, we select informative negative items from the $C$ previous candidates in a similar manner as in \cite{wu:iccv17}. 
Given the latent vector of a positive item $\mathbf{v}_{j}$, we sample a negative item index $n$, with corresponding latent factor $\mathbf{v}_{n}$ as follows:
\begin{displaymath}
    \Pr(n|\mathbf{v}_{j}) \propto \begin{cases}
    \frac{1}{p(\mathbf{v}_{j}^{T}\mathbf{v}_{n}=s)}, & 0 \leq s \leq 1 \\
    0, & \text{otherwise}
    \end{cases}
\end{displaymath}
This strategy has two objectives: first, the choice of this probability function offers triplets with a larger range of gradient's variance than what would be obtained with semi-hard triplet sampling \cite{wu:iccv17}. Second, it puts high probability on items $n$ that produce high positive value for $\mathbf{v}_{j}^{T}\mathbf{v}_{n}$, hence inducing positive values for $L^{\text{triplet}}$ and large values for $L^{\text{GOR}}$. Indeed, it's obvious that with positive $\mathbf{v}_{j}^{T}\mathbf{v}_{n}$, $L^{\text{GOR}}$ increases as $\mathbf{v}_{j}^{T}\mathbf{v}_{n}$ gets higher. At the same time, for each positive-negative pair $(\mathbf{v}_{j}, \mathbf{v}_{n})$, we have $||\mathbf{v}_{j} - \mathbf{v}_{n}||_{2}^{2} = ||\mathbf{v}_{j}||_{2}^{2} + ||\mathbf{v}_{n}||_{2}^{2} - 2\mathbf{v}_{j}^{T}\mathbf{v}_{n} = 2 - 2\mathbf{v}_{j}^{T}\mathbf{v}_{n}$, so the greater the value of $\mathbf{v}_{j}^{T}\mathbf{v}_{n}$ is, the closer the positive-negative points are. This leads to a smaller difference between $D^2(\mathbf{u}_{i}, \mathbf{v}_{j})$ and $D^2(\mathbf{u}_{i}, \mathbf{v}_{n})$, making $L^{\text{triplet}}$ more likely to be positive. It thus induces higher loss values compared to the uniform sampling case, and hopefully results in gradients more suitable for training.
\section{Experiments}

\subsection{Experimental Settings}
\subsubsection{Datasets}
We experiment with three datasets covering different domains: namely movie, book and music recommendations.

\textit{Amazon movies} \cite{he:www16}: The \textit{Amazon} dataset is the consumption records with reviews from \textit{Amazon.com}. We use the user-movie rating from \textit{the movies and tv category 5-core}. The data is binarized by keeping only ratings greater than 4 as implicit feedback. Users with less than 20 positive interactions are filtered out.

\textit{Book crossing} \cite{ziegler:www05}: The dataset contains book ratings which scale from 0 to 10 with the higher score indicating preference. Again, explicit ratings are binarized by keeping values of five or higher as implicit feedback. Only users with more than 10 interactions are then kept.

\textit{Echonest} \cite{mcfee:www12}: The EchoNest Taste Profile dataset contains user playcounts for songs of the \ac{MSD}. After deduplicating songs, playcount data is binarized by considering values of five or higher as implicit feedback. Finally, only users with more than 20 interactions and items with which at least 5 users interacted.

The characteristics of these three datasets after filtering are summarized in Table~\ref{tab:commands}.

\subsubsection{Evaluation Methodology}
We divide user interactions into 4-fold for cross-validation where three folds are used to train the model and the remaining fold is used for testing. Based on the ranked preference scores, we adopt \ac{MAP} and \ac{NDCG} to measure whether ground-truth items are present on the ranked list of preferences truncated at 50 items and their positions. In addition, we calculate the \ac{MMR} of recommended items to assess the popularity bias of the model.

\subsubsection{Parameters setting}
The parameter $C$ should be chosen in order to retain a sufficient number of candidates while limiting the amount of computations occurring in the second stage. We set it to 2000 and leave its optimization to future work. Besides that, the latent dimension $d$ is set to 128 and the margin $\alpha$ to 1.
For the other parameters, the 4-fold cross-validation mentioned above is used to choose the best values using grid-search. Adam optimizer \cite{kingma:arxiv14} is used for all models. The learning rate is 0.0001, the parameter $\beta$ for the first stage is 1.0 for Amazon movies and Echonest and 0.8 for Book crossing. Finally, $\lambda_g$ is 0.01 when the number of negatives is 1 and 2 and to 0.001 as the number of negatives is 5.

\begin{table}
  \caption{Statistics of the datasets}
  \label{tab:commands}
  \begin{tabular}{ccccl}
    \toprule
    Dataset & User\# &Item\# & Rating\# & Density \\
    \midrule
    \texttt{Amazon Movies} & 11181 & 94661 & 620059 & 0.058\% \\
    \texttt{Book Crossing} & 3593 & 127339 & 240020 & 0.052\% \\
    \texttt{Echonest} & 31521 & 159063 & 1405671 & 0.028\% \\
    \bottomrule
  \end{tabular}
  \vspace{-4mm}
\end{table}

\subsection{Comparison Results}
\subsubsection{Uniform sampling}
Performance of \ac{CML} with uniform sampling \cite{hsieh:www17} is summarized in Table \ref{all_results} (\textit{Uni} sub-table). We discuss results for the Amazon movies dataset as the same trend can be observed on the two others. We see that the performance of \ac{CML} in terms of MAP and NDCG heavily decreases when using small batches, especially when $|N_{ij}|=1$. For example, when the batch size is an order of magnitude smaller (256 vs 4096), \ac{MAP} relatively decreases by 19\% (2.26 $\rightarrow$ 1.82) and NDCG by 14\% (7.55 $\rightarrow$ 6.47). This drop supports the idea that the number of informative triplets is low in small batches with the uniform sampling setup. With more negatives per triplet ($|N_{ij}|=5$), this decrease is alleviated, about 7\% relative drop against 19\% for \ac{MAP} (2.48 $\rightarrow$ 2.31) and 5\% relative drop against 14\% for NDCG (8.06 $\rightarrow$ 7.68). Additionally, another issue of \ac{CML} is being prone to a strong popularity bias (\ac{MMR}). As shown in Table \ref{all_results} this bias increases with the batch size: e.g., from 256 to 4096, with 1 negative per triplet, MMR raises relatively by 29\% (86.4 $\rightarrow$ 111.8).

\begin{table*}[h]
  \caption{\ac{CML}'s performance with different sampling strategies, number of negatives per triplet and batch sizes. The format is $mean \pm std$ obtained from 4 runs on cross-validation splits. The italic bold face shows the best values for uniform strategy while bold face shows the best overall values}.
  \label{all_results}
  \resizebox{\textwidth}{!}{
  \begin{tabular}{|c|c|c|ccc|ccc|ccc|}
        \toprule
        \multirow{2}{*}{Sam} & \multirow{2}{*}{$|N_{ij}|$} & \multirow{2}{*}{Batch} & \multicolumn{3}{c}{Amazon Movies} & \multicolumn{3}{c}{Book Crossing} & \multicolumn{3}{c}{Echonest} \\
        \cline{4-12}
        {} & {} & {} & MAP(\%) & NDCG(\%) & MMR & MAP(\%) & NDCG(\%) & MMR & MAP(\%) & NDCG(\%) & MMR \\
        \midrule
        \multirow{6}{*}{Uni} & \multirow{3}{*}{1} & {4096} & {2.26 $\pm$ 0.06} & {7.55 $\pm$ 0.12} & {111.8 $\pm$ 0.1} & {1.12 $\pm$ 0.07} & {3.98 $\pm$ 0.11} & {55.2 $\pm$ 0.4} & {4.88 $\pm$ 0.02} & {12.78 $\pm$ 0.01} & {241.8 $\pm$ 1.4} \\
        {} & {} & {1024} & {2.13 $\pm$ 0.08} & {7.19 $\pm$ 0.16} & {101.7 $\pm$ 1.5} & {1.08 $\pm$ 0.08} & {3.86 $\pm$ 0.14} & {53.2 $\pm$ 0.4} & {4.41 $\pm$ 0.01} & {11.70 $\pm$ 0.04} & {196.5 $\pm$ 0.6} \\
        {} & {} & {256} & {1.82 $\pm$ 0.04} & {6.47 $\pm$ 0.10} & {\textbf{\textit{86.4 $\pm$ 2.4}}} & {0.93 $\pm$ 0.04} & {3.50 $\pm$ 0.06} & {\textbf{\textit{44.5 $\pm$ 1.2}}} & {3.66 $\pm$ 0.03} & {9.90 $\pm$ 0.12} & {\textbf{\textit{146.7 $\pm$ 2.2}}} \\
        \cline{2-12}
        {} & \multirow{3}{*}{2} & {4096} & {2.34 $\pm$ 0.06} & {7.72 $\pm$ 0.10} & {114.5 $\pm$ 0.2} & {1.15 $\pm$ 0.08} & {4.06 $\pm$ 0.12} & {54.9 $\pm$ 0.9} & {5.16 $\pm$ 0.04} & {13.39 $\pm$ 0.05} & {265.5 $\pm$ 1.1} \\
        {} & {} & {1024} & {2.23 $\pm$ 0.04} & {7.49 $\pm$ 0.10} & {109.5 $\pm$ 0.2} & {1.13 $\pm$ 0.06} & {3.97 $\pm$ 0.13} & {54.5 $\pm$ 1.2} & {4.85 $\pm$ 0.08} & {12.60 $\pm$ 0.09} & {231.7 $\pm$ 3.0} \\
        {} & {} &  {256} & {2.04 $\pm$ 0.07} & {6.98 $\pm$ 0.12} & {96.0 $\pm$ 1.9} & {1.04 $\pm$ 0.10} & {3.74 $\pm$ 0.17} & {49.5 $\pm$ 0.8} & {4.18 $\pm$ 0.06} & {11.14 $\pm$ 0.09} & {175.0 $\pm$ 5.3} \\
        \cline{2-12}
        {} & \multirow{3}{*}{5} & {4096} & {\textbf{\textit{2.48 $\pm$ 0.05}}} & {\textbf{\textit{8.06 $\pm$ 0.13}}} & {118.5 $\pm$ 0.4} & {\textbf{\textit{1.15 $\pm$ 0.08}}} & {\textbf{\textit{4.13 $\pm$ 0.14}}} & {53.7 $\pm$ 0.8} & {\textbf{\textit{5.71 $\pm$ 0.04}}} & {\textbf{\textit{14.50 $\pm$ 0.09}}} & {315.5 $\pm$ 1.4} \\
        {} & {} &  {1024} & {2.45 $\pm$ 0.06} & {8.01 $\pm$ 0.13} & {115.2 $\pm$ 0.6} & {1.15 $\pm$ 0.08} & {4.07 $\pm$ 0.09} & {55.2 $\pm$ 0.6} & {5.56 $\pm$ 0.05} & {14.17 $\pm$ 0.04} & {287.5 $\pm$ 2.5} \\
        {} & {} &  {256} & {2.31 $\pm$ 0.02} & {7.68 $\pm$ 0.04} & {107.2 $\pm$ 0.2} & {1.12 $\pm$ 0.09} & {3.94 $\pm$ 0.16} & {53.5 $\pm$ 1.0} & {5.11 $\pm$ 0.01} & {13.10 $\pm$ 0.04} & {229.1 $\pm$ 1.0} \\
        \hlinewd{0.9pt}
        \multirow{3}{*}{Pop} & {1} & \multirow{3}{*}{256} & {2.12 $\pm$ 0.02} & {7.14 $\pm$ 0.01} & {26.2 $\pm$ 0.3} & {1.04 $\pm$ 0.07} & {3.54 $\pm$ 0.06} & {8.3 $\pm$ 0.04} & {6.48 $\pm$ 0.11} & {15.55 $\pm$ 0.17} & {54.3 $\pm$ 1.6} \\
        {} & {2} & {} & {2.43 $\pm$ 0.04} & {7.83 $\pm$ 0.12} & {24.3 $\pm$ 0.02} & {1.22 $\pm$ 0.07} & {3.87 $\pm$ 0.14} & {6.9 $\pm$ 0.13} & {7.26 $\pm$ 0.06} & {17.0 $\pm$ 0.03} & {46.1 $\pm$ 0.2} \\
        {} & {5} & {} & {2.57 $\pm$ 0.07} & {7.89 $\pm$ 0.06} & {\textbf{17.2 $\pm$ 0.04}} & {1.25 $\pm$ 0.04} & {3.75 $\pm$ 0.06} & \textbf{{4.1 $\pm$ 0.01}} & {8.0 $\pm$ 0.01} & {18.44 $\pm$ 0.08} & {\textbf{33.8 $\pm$ 0.3}} \\
        \hlinewd{0.9pt}
        \multirow{3}{*}{2st} & {1} & \multirow{3}{*}{256} & {2.32 $\pm$ 0.06} & {7.78 $\pm$ 0.14} & {32.9 $\pm$ 0.1} & {1.26 $\pm$ 0.08} & {\textbf{4.13 $\pm$ 0.13}} & {10.1 $\pm$ 0.1} & {6.99 $\pm$ 0.03} & {16.69 $\pm$ 0.05} & {96.1 $\pm$ 0.2} \\
        {} & {2} & {} & {2.57 $\pm$ 0.03} & {8.14 $\pm$ 0.04} & {27.9 $\pm$ 0.1} & {1.3 $\pm$ 0.08} & {4.12 $\pm$ 0.11} & {7.6 $\pm$ 0.1} & {7.91 $\pm$ 0.14} & {18.32 $\pm$ 0.13} & {78.9 $\pm$ 0.2} \\
        {} & {5} & {} & {\textbf{2.73 $\pm$ 0.03}} & {\textbf{8.23 $\pm$ 0.02}} & {19.6 $\pm$ 0.1} & {\textbf{1.38 $\pm$ 0.09}} & {4.12 $\pm$ 0.15} & {4.5 $\pm$ 0.01} & {\textbf{8.70 $\pm$ 0.07}} & {\textbf{19.37 $\pm$ 0.06}} & {48.4 $\pm$ 0.8} \\
        \bottomrule
      \end{tabular}
    }
\vspace{-2mm}
\end{table*}
\subsubsection{Popularity-based sampling}
To confirm our intuition on the necessity of pushing non-matching popular items farther away (as discussed in the Section \ref{2_stage_negative_sampling}), we study the popularity-based negative sampling method of Equation \eqref{w2vsampling}. Table \ref{all_results} (\textit{Pop} sub-table) reveals a high impact of this strategy on the performance of \ac{CML} in terms of \ac{MAP} \& \ac{NDCG}. Specifically, with smaller batch size (256 vs 4096), the \ac{MAP} with popularity-based sampling already surpasses the best result of the uniform sampling, by 3.6\% for movies (2.26 $\rightarrow$ 2.57), 8.7\% for books (1.15 $\rightarrow$ 1.25) and 40\% for music (5.71 $\rightarrow$ 8.0) respectively. As expected, the recommendations are less biased towards popular items: \ac{MMR} decreases by 80\%  on Amazon movies (86.4 $\rightarrow$ 17.2), 90.8\% on Book crossing (44.5 $\rightarrow$ 4.1) and 77\% on Echonest (146.7 $\rightarrow$ 33.8). 
\subsubsection{2-stage sampling}
 While popular-based negative sampling is efficient with small batches, the reported \ac{NDCG} of this strategy is slightly worse than what can be obtained by uniform sampling on large batches (except for the Echonest). In book recommendation, the gap is quite significant with a 6.3\% decrease (4.13 $\rightarrow$ 3.87). To further improve the performances of the \ac{CML} model on small batches, we add on top of the popularity strategy a second stage based on dot product weighted sampling as described in Section \ref{2_stage_negative_sampling}. This enables \ac{CML} with small batches to have a competitive \ac{NDCG} w.r.t the best result using the uniform sampling strategy. In detail, with 16 times smaller batch size, 2-stage sampling yields the same \ac{NDCG} for book recommendation and reaches a 2.1\% increase for movie recommendation (8.06 $\rightarrow$ 8.23), 33.6\% increase for music recommendation (14.5 $\rightarrow$ 19.37). Meanwhile \ac{MAP} is remarkably enhanced for all datasets, 10\% (2.48 $\rightarrow$ 2.73), 20\% (1.15 $\rightarrow$ 1.38) and 52\% (5.71 $\rightarrow$ 8.7) for movie, book and music recommendation respectively. Note that 2-stage sampling makes the \ac{MMR} slightly higher than that of the popularity-based strategy, but it is still significantly lower than the one with uniform sampling.

\section{Conclusions}
We proposed a 2-stage sampling strategy that enables the \ac{CML} model to perform efficiently with batch size an order of magnitude smaller than what would be needed with the default uniform sampling. At its heart, a set of samples is first selected based on their popularity. Then, informative ones are drawn from this set based on their inner product weights with anchors. Experiments demonstrate positive results across various datasets, especially for music recommendation for which the proposed approach increased very significantly the performance of the system. In future work, we will leverage this sampling strategy to jointly learn from multimedia content and collaborative data where huge batches are prohibitive due to memory limitations.

\bibliographystyle{ACM-Reference-Format}
\bibliography{sample-bibliography}

\end{document}